# Modelization of RGB lasers based on QPM structures with independent control of laser intensities by electrooptic effect

R. Vilaplana and F. J. Manjón[*]

Dpto. de Física Aplicada, Universitat Politècnica de València, E.P.S.A.
E-03801 Alcoy (Spain)

**Abstract:** Recently simultaneous generation of continuous-wave visible laser light in the three fundamental colors has been experimentally demonstrated in $Nd^{3+}$-doped $LiNbO_3$ and $Sr_{0.6}Ba_{0.4}(NbO_3)_2$ by quasi-phase matching intracavity self-frequency conversion. In this work, we report the modelization of RGB lasers operating under similar experimental conditions but designed in the simplest quasi-periodically poled ferroelectric form by using the Fourier Transform technique. The aim of this work is to analyse the possibility of controlling the three laser intensities in an independent and novel way by means of the linear electrooptic (Pockels) effect in these non-linear materials. Our results suggest that the application of electrooptic effect for this purpose, according to our proposed scheme, is better adapted to $Sr_{0.6}Ba_{0.4}(NbO_3)_2$ than to $Nd^{3+}$-doped $LiNbO_3$.

**PACS numbers:** 42.55. Rz, 42.55. Xi, 42.60.By, 42.65.Ky

**Key words**: PPLN, APLN, $LiNbO_3$, SBN, $Sr_xBa_{1-x}(NbO_3)_2$, solid-state-lasers, electrooptic effect

---

[*] Corresponding author to whom correspondence should be addressed.
  e-mail: fjmanjon@fis.upv.es, Tel.:+34 96 652 84 42, Fax: +34 96 652 84 09

Simultaneous generation of continuos-wave visible laser radiation in the three fundamental colors red, green, and blue (RGB) has potential applications in laser-light-based high-brightness displays [1], and considerable attention has been focused on RGB lasers during the last few years [2-8]. In this sense, simultaneous laser light in the three fundamental colors has been recently demonstrated using the intracavity self-frequency conversion (SFC) of infrared laser light of $Nd^{3+}$ ions operating under quasi-phase matching (QPM) conditions in aperiodically-poled $Nd^{3+}$-doped $LiNbO_3$ (Nd:APLN) [6] and also in $Sr_{0.6}Ba_{0.4}(NbO_3)_2$ pumped with $Nd^{3+}$-doped $YVO_4$ [7].

The study of SFC laser emission in rare-earth doped optical superlattice crystals had just begun and is becoming increasingly more interesting because of the ability of these crystals to generate multicolor lasers simultaneously across the visible part of the spectrum through a single crystal. SFC lasers provide a way of obtaining visible radiation from efficient laser ions in the near infrared. In particular, red, green, and blue laser radiation can be originated by self-frequency doubling (SFD) or self sum-frequency mixing (SSFM) of the pump radiation and the two IR laser emissions from the $^4F_{3/2} \rightarrow {}^4I_{11/2}$ and $^4F_{3/2} \rightarrow {}^4I_{13/2}$ $Nd^{3+}$ crystal-field (CF) transitions. The $^4F_{3/2} \rightarrow {}^4I_{11/2}$ and $^4F_{3/2} \rightarrow {}^4I_{13/2}$ CF transitions are located around 1 and 1.3 μm respectively, and the laser pump radiation is typically in the 0.7-0.8 μm region.

In Nd:APLN, IR laser emissions of $Nd^{3+}$ ions were obtained by pumping with laser at 744 nm. Red (686 nm) and green (542 nm) laser emissions were obtained by SFD and blue (441 nm) was obtained by SSFM [6]. Several other visible laser emissions were observed in this system likely due to random errors in periodic poling, as suggested by Chen et al. [8]. On the other hand, IR laser emissions of $Nd^{3+}$ ions were focused onto $Sr_{0.6}Ba_{0.4}(NbO_3)_2$, hereafter noted as SBN60, to obtain RGB laser emission [7]. IR laser emissions from $Nd^{3+}$ ions were produced in a $Nd^{3+}$-doped $YVO_4$ crystal pumped at 880 nm. Red (671 nm) and blue (440 nm) laser emissions were obtained by SFD and green (531 nm) by SSFM inside the SBN60 compound. Fig. 1 shows the pump and infrared $Nd^{3+}$ laser emissions used for simultaneous RGB emission in both the Nd:APLN and Nd:$YVO_4$-SBN60 systems.

Following Chen et al. [8], we think that RGB laser action in the two mentioned compounds [6,7] are likely due to uncontrolled errors in periodic poling resulting in broad tuning curves. In this work, we propose a model of regular structures of ferroelectric poling to get RGB laser emissions using the QPM method in a similar way to those experimentally observed. The model is based on the simplest regular quasiperiodic structures of ferroelectric domains that can be performed in $LiNbO_3$ (LN) and SBN. In our quasiperiodic structure we separate the contribution of each crystal zone to the amplification of each color. In this way, we explore the possibility of controlling the RBG laser intensities in an independent way by means of the first-order electrooptic (Pockels) effect. For that purpose, we study the work of the display shown in Fig. 2, as already suggested in Ref. [9].

## 1. QPM design principles and results

QPM is a technique for phase matching non-linear optical interactions in which the relative phase is corrected at regular intervals using a structural periodicity built in the non-linear medium [10,11]. This technique is based in the idea that efficient non-linear interactions require compensation for the phase mismatch between the electromagnetic waves, due to the different phase velocities of the waves caused by the dispersion of the non-linear material. This phase difference produces a π phase shift between the output wave and the interacting or pump waves over every certain distance, named coherence length $l_c$, which leads to a reversal of the energy flow from the output wave to the pump waves. In non-linear frequency conversion; e.g., to obtain a laser emission from two interacting waves 1 and 2, both the energy law conservation and the momentum law conservation must be obeyed

$$E_{laser} = E_1 + E_2 \tag{1}$$

$$E_{laser} \cdot n_{laser}(E_{laser}) = E_1 \cdot n_1(E_1) + E_2 \cdot n_2(E_2) \tag{2}$$

The idea of quasi-phase matching (QPM) is that Eq. (2) can be satisfied along the whole crystal by introducing a periodic change in the sign of the medium's non-linear susceptibility along one axis of the non-linear crystal that resets the phase of the polarisation wave by π, with half-period equal to the coherence length $l_c$. In this way, the power can be made to flow continuously from the pump waves into the output wave.

The QPM condition can be achieved in ferroelectric materials by introducing a one-dimensional periodic structure that reverse the domain orientations, and consequently reverse the sign of the non-linear susceptibility. The spatial periodic structure contains a series of reciprocal vectors, being integer times of the primitive one, and the combination of one reciprocal vector with the wave vector of the interacting beam results in the QPM condition. The non-linear optical interactions using the QPM method have smaller conversion efficiencies than those using conventional birefringent phase matching methods. This drawback is compensated by placing the non-linear medium inside a laser cavity to achieve intracavity conversion and coupling the pump and output waves polarised parallel to each other along the diagonal components of the non-linear susceptibility tensor, which is often the largest one of the crystal.

The first-order QPM condition can be written as:

$$\Delta k \cdot l_c = \pi \tag{3}$$

where $\Delta k$ is the phase mismatch wavevector between the interacting and the output waves and $l_c$ is the coherence length. If the laser emission is obtained by a SSFM process the coherence length for the laser emission is given by

$$l_c = \frac{hc/2}{\left[E_{laser} \cdot n(E_{laser}) - E_1 \cdot n(E_1) - E_2 \cdot n(E_2)\right]} \tag{4}$$

where $E_{laser}$ and $n(E_{laser})$ are the energy and refractive index of the SSFM emission respectively, and subindices 1 and 2 refer to the two interacting waves, respectively. The coherence length for a laser emission given by a SFD process can be readily obtained making $E_1 = E_2$ and $n(E_1) = n(E_2)$ in Eq. (4). In this way, the QPM method allows to phase match a single interaction by periodic variation of the ferroelectric domains alternating positive domains of length $l_p = l_c$ and negative domains of length $l_n = l_c$.

The curve representing the efficiency of a non-linear process as a function of the phase shift of the interacting waves (tuning curve) is related to the normalized spatial distribution of ferroelectric domains in the crystal through the Fourier transform (FT) [11]. This view turns out to be mathemathically convenient for computational tasks. In this sense, the short Fourier transform (STFT) or the Wigner transform can give a simultaneous representation of the total length of the crystal and the phase mismatch wavevector. We have calculated the contribution of three separated ferroelectric domain structures to the amplification of the three wavelengths experimentally obtained in Nd:APLN [6] and in SBN60 [7], according to the schematic model shown in Fig. 2.

Taking into account the extraordinary refractive index dispersion of LiNbO$_3$ [12], we obtain that ferroelectric domains with coherence length of 1.60 μm for blue, 3.64 μm for green, and 6.93 μm for red laser emissions should be present in our quasi-periodic structure of Nd-doped LN to obtain the same RGB laser emissions as in Nd:APLN [6]. In analogous way, considering the extraordinary refractive index dispersion of SBN60 [13], ferroelectric domains with coherence lengths of 2.11 μm (blue), 2.70 μm (green), and 5.44 μm (red) should be present in SBN60 to obtain the emissions experimentally reported [7].

A major drawback for the control of the RGB laser emission with the electrooptic effect is that there is no net electrooptic effect along the crystal if the positive and negative domains are of equal length (50%-duty cycle) [14]. It is necessary to build up a positive domain with length $l_p = x \cdot \Lambda$ different to that of the negative domain $l_n = (1-x) \cdot \Lambda$, being x the asymmetry parameter and $\Lambda = 2 \cdot l_c$. This modification of the domain lengths affects the QPM condition (Eq. (3)) by modifying the overlapping of the QPM peak and the laser peak corresponding to SFD or SSFM processes. A consequence of this overlap mismatch is that the laser conversion efficiencies in presence of the asymmetry x are lower than those obtained for the symmetric case. The laser conversion efficiencies can be obtained by calculating the Fourier transform.

Inset of Fig. 2 shows the periodic repetition of micrometric structures of length $\Lambda = 2 \cdot l_c$. Each micrometric structure consists of one positive and one negative ferroelectric domain with different length. For calculation of the non-linear process efficiency, the ferroelectric domains have been modelled with a quasi-periodic finite square signal oscillating between the 1 (positive domain) and −1 (negative domain) values. The number of points 1 or −1 inside each finite domain is related to the lengths of the positive and negative domains, respectively. In this way, the FT of the amplitude of the interacting waves versus the length of the crystal gives us the non-linear conversion efficiency as a function of the phase mismatch wavevector. If we apply the STFT or the Wigner transform we get the non-linear conversion efficiency as a function of the representation of the phase mismatch wavevector and of the crystal length at the same time. This latter transform allows us to analyse which part of the crystal is contributing to the amplification of each laser peak since each laser peak is amplified for a given phase mismatch wavevector. Fig. 3 shows the non-linear conversion efficiency as a function of the phase mismatch wavevector for Nd-doped quasi-periodically poled $LiNbO_3$ and the correlation between the zone of the crystal contributing to each phase mismatch wavevector (each laser light). For the calculations, we have taken a LN crystal 0.5-mm long with three periodic structures. In order to obtain the same efficiency for the generation of the three colors the FT imposes that the length of the total region contributing to each color must be the same. This means that the 1-mm crystal has been divided into three equal parts each one contributing to the amplification of one single color. As the coherent length of the blue color is smaller than that of the green and red colors, this means that for the same crystal length to amplify each color the number of quasi-periodic micrometric structures contributing to the amplification of each color increases when going from blue to red.

As previously noted, the use of quasi-periodic structures in non-linear crystals due to asymmetry leads to the reduction of the non-linear process efficiencies. Fig. 4 shows the decrease of the efficiency of the non-linear SFD and SSFM processes in the RGB region for LN as a function of the asymmetry parameter. We can observe that for asymmetries between 20 and 30% the non-linear conversion efficiencies can be still large enough for practical applications.

## 2. Electrooptic design principles and results

Recent reports demonstrate the possibility of frequency tuning by means of electrooptic effect in LiNbO$_3$ and PPLN [14-17]. However, to our knowledge, the control of laser intensity by means of linear electrooptic effect has been only considered by Tran and Furlan [18] and by H. Ridderbusch et al. [19,20]. In these works, the control of the intensity of the generated laser was performed by the phase mismatch induced in either a longitudinal or a transverse Pockels cell. Instead, we consider the possibility of using the first-order electrooptic (Pockels) effect to control the intensity of the RGB laser lights directly inside of the non-linear active medium without phase mismatch. This can be done by shifting the QPM condition (QPM peak) from the laser peak in Eq. (3). The QPM peak shift can be done by changing the extraordinary refractive index with the application of an electric field along the optical axis making use of the large electrooptic coefficients r$_{33}$ along the *z* axis of the LN, and especially of the SBN compounds.

Under applied electric field, the first-order QPM condition given by Eq. (3) can be written as

$$\Delta k_p \cdot l_p + \Delta k_n \cdot l_n = 2\pi \quad (5)$$

with $l_p$ and $l_n$ the positive and negative ferroelectric domains, respectively. $\Delta k_p$ and $\Delta k_n$ are the phase mismatch wavevectors between the QPM peak and the two input interacting waves in the positive and negative domains, respectively. Since the phase mismatch wavevector is modified due to the modification of the refractive index by the applied electric field, Eq. (5) for the SFMM process can be written as

$$[E_{QPM} n_+(E_{QPM}) - E_1 n_+(E_1) - E_2 n_+(E_2)] \cdot l_p +$$
$$+ [E_{QPM} n_-(E_{QPM}) - E_1 n_-(E_1) - E_2 n_-(E_2)] \cdot l_n = hc \qquad (6)$$

where $E_{QPM}$ is the new energy satisfying the QPM condition if there is an applied electric field, and $n_+$ and $n_-$ are the modified refractive indices in the positive and negative ferroelectric domains, respectively. Again, the new QPM condition with applied electric field for SFD can be readily obtained making $E_1 = E_2$, $n_+(E_1) = n_+(E_2)$, and $n_-(E_1) = n_-(E_2)$.

In order to apply Eq. (6) we have to consider that LiNbO$_3$ crystallises in the trigonal structure and belongs to the 3m point group symmetry; therefore, the compound is characterised by the electrooptic constants $r_{22}$, $r_{13}$, $r_{33}$, and $r_{51}$. On the other hand, Sr$_x$Ba$_{1-x}$Nb$_2$O$_6$ compounds with 0.25<x<0.75 crystallise in the tetragonal tungsten bronze structure belonging to the 4mm point group symmetry [21]. These compounds are characterised by the electrooptic constants $r_{31}$, $r_{33}$, and $r_{51}$ [22]. Both LN and SBN are uniaxial crystals and the crystal remains uniaxial with the same principal axis when an electric field is applied along the optic axis (z axis). However, the refractive indices along the principal axis are slightly modified [23]. In this sense, the extraordinary refractive index in presence of an applied electric field **E** in the positive and negative ferroelectric domains can be written as [24]

$$n_{e\pm}(\mathbf{E}) = n_e(0) \mp \frac{1}{2} n_e^3(0) r_{33} \mathbf{E} \qquad (7)$$

The change of the refractive index leads to the modification the QPM condition of Eq. (3) and the QPM condition of Eq. (6) will be now satisfied by another energy $E_{QPM} \neq E_{laser}$, where $E_{laser}$ is the frequency satisfying the QPM condition in Eq. (3); i.e., in absence of the applied electric field, **E**.

Considering $l_p = x \cdot \Lambda$ and $l_n = (1-x) \cdot \Lambda$, with $\Lambda = 2 \cdot l_c$, we can obtain from Eqs. (6) and (7) the electric field necessary to shift the QPM condition away from the laser frequency for SFD or SSFM processes as a function of the asymmetry parameter. For the SSFM process, we obtain

$$\mathbf{E} = \frac{1}{r_{33}} \frac{\frac{2hc}{\Lambda} - 2E_{QPM}\, n(E_{QPM}) + 2E_1\, n(E_1) + 2E_2\, n(E_2)}{(1-2x)\left[E_{QPM}\, n^3(E_{QPM}) - E_1\, n^3(E_1) - E_2\, n^3(E_2)\right]} = \\ = \frac{2}{r_{33}} \frac{E_{laser}\, n(E_{laser}) - E_{QPM}\, n(E_{QPM})}{(1-2x)\left[E_{QPM}\, n^3(E_{QPM}) - E_1\, n^3(E_1) - E_2\, n^3(E_2)\right]}$$

(8)

where $r_{33}$ is the electrooptic coefficient along the c axis, and $E_{QPM}$ and $n(E_{QPM})$ are the energy of the QPM peak, satisfying the QPM condition with applied electric field, and its extraordinary refractive index. On the other hand, $E_{laser}$ and $n(E_{laser})$ are the laser peak energy, due to SFD or SSFM processes, and its extraordinary refractive index, and similarly for the two input interacting waves 1 and 2. Eq. (8) gives the electric field necessary to shift the QPM peak from the laser peak as a function of the energy difference between the laser and QPM peaks and as a function of the asymmetry parameter. The electric field **E** needed to shift the QPM peak from the laser peak for the SFD process can be readily obtained by making $E_1 = E_2$ and $n(E_1) = n(E_2)$ in Eq.(8).

Fig. 5 shows how the electric field works. In absence of electric field [Fig. 5(a)] the QPM condition is almost totally satisfied, except for the small asymmetry, and the QPM peak frequency almost coincides with the peak of the laser emission. In this case, we obtain the maximum laser intensity since the color intensity is related to the convolution of the laser peak and the QPM peak. If we apply a small electric field [Fig. 5(b)] the QPM peak is shifted from the laser peak thus reducing the efficiency of the emission. Finally, if the electric field is large enough [Fig. 5(c)] we can shift the QPM peak far away from the laser peak, thus obtaining no emission for the color of the laser peak desired. In this way and making use of the design of Fig. 2 it is possible to control the three RGB laser intensities in an independent way by applying different electric fields to each different region contributing to a single color.

For technical applications of Eq. (8) we have to make several considerations. Equation (8) indicates that the largest the asymmetry parameter x the lowest the electric field needed to shift the QPM peak from the laser peak. Low values of the electric fields are key to effective control of laser lights since values obtained from Eq. (8) are rather high. However, since the conversion efficiency of the non-linear process decreases with the asymmetry (see Fig. 4), we must take a compromise between the conversion efficiency of the non-linear process and the electric field necessary for the effective control of the laser intensities. Fig. 6 shows the electric fields needed to shift the QPM peak by 0.2 nm away from the red, green, and blue laser peaks in SBN60 ($r_{33} \approx 420$ pm/V [25]) as a function of the asymmetry parameter x according to Eq. (8). Only values up to x = 0.5 (totally symmetric) have been plotted in Fig. 6 since values above x = 0.5 are similar to those obtained below 0.5 but with the sign of the electric field changed (reverse bias). Fig. 6 also shows the calculated values required to shift the QPM peak by 0.2 nm for the three colors in SBN75 ($r_{33} \approx 1380$ pm/V [26]) if we use the refractive index of SBN60 [13] and assuming the same frequencies of the three fundamental colors observed in SBN60 [7].

Just for comparison, we give the electric field values needed to shift the QPM peak by 0.2 nm from the laser peaks in LN ($r_{33} \approx 30.8$ pm/V [21]) in a 25%-75% duty cycle (x = 0.25) and with the main visible laser frequencies observed in Nd:APLN [6]. In such case, we obtain electric field values of 90 kV/mm (for 441 nm), 110 kV/mm (for 542 nm), and 130 kV/mm (for 686 nm) respectively. These values are by far larger than those used for frequency tuning in optical parametric oscillation (OPO) in LN [17] and much larger than the coercive field of the compound at room temperature (21 kV/mm). The different values found in SFD or SSFM processes as compared to OPO processes are due to the fact that in SFD and SSFM there are two input interacting waves generating a third one while in OPO there is only one input wave that generates two waves. Therefore, in OPO processes there is a compensation between the shifts of the signal and idler waves with the applied electric field and frequency tuning of both signal and idler takes place at moderate voltages. However, the separation of the QPM peak respect to the laser emission in SFD and SSFM processes is not compensated and the electric fields needed to tune the non-linear conversion efficiency are much larger in these processes.

In the case of $LiNbO_3$ the voltages needed to control the laser intensity with the scheme proposed in Fig. 2 are far beyond the breakdown field in air (~1 kV/mm [15]) and beyond the coercive field at room temperature (21 kV/mm), so the application of the linear electrooptic effect for this purpose seems to be not possible. However, in SBN and related materials the situation could be rather different, since a recent study evidences that the breakdown field in SBN increases with the increase of the Sr content [27]. In this sense, the coercive field has been measured between 0.13 and 1 kV/mm in SBN60 [28], and the breakdown field seems to be around 4 kV/mm [27]. These values are near the values calculated to shift the QPM peak by 0.2 nm from the laser peak. Therefore, this material could be used for controlling laser intensity by electrooptic effect according to the scheme shown in Fig. 2. The situation could be even better for SBN75, since the voltages needed to control laser intensity are much lower than the coercive field (>0.5 kV/mm [29]) and breakdown field (>5.7 kV/mm [27]), so SBN75 could be a good candidate for high brightness displays controlled by electrooptic effect according to our design principle.

The maximum miniaturization of the RGB laser display could be achieved if Nd doping of the SBN is accomplished, as recently proved by [30,31]. This doping seems to lead to an even a further increase of the electrooptic coefficient as well as to an increase of the conductivity that should decrease the breakdown field. We think that the possibilities of controlling laser lights in an independent way in these systems are worth to be explored. We have to note that in this work we have not analysed the effect of the electric field on the Stark levels of $Nd^{3+}$ ions inside the solid-state laser host. We know that the effect of the applied electric field could also affect considerably the energies of the $Nd^{3+}$ Stark levels ($^4F_{3/2}$, $^4I_{11/2}$, and $^4I_{13/2}$) involved in the IR laser emissions and, consequently, affect the RGB laser frequencies and intensities and the electric fields needed for the control of the RGB laser intensities. In this sense, the application of electric field to control laser intensities in systems with different Nd concentrations are also worth to be explored.

Finally, we want to consider several limitations of the design proposed in Fig. 2. The application of this design principle to control laser intensities in an independent way can be greatly hindered in practice by periodic poling errors (linearly tapered period errors and random period errors). In this sense, periodic poling errors usually lead to broad tuning curves [8], so much larger electric fields will be needed to shift the centre of the QPM peak from the laser peak. In this sense, random period errors seem to be more undesirable than tapered period errors because the latter give a finite broadening of the QPM peak in the tuning curve while the former give a large broadening which could avoid the effective control of the zero intensity level. On the other hand, Eq. (8) yields larger electric fields for larger deviations of the QPM peak with respect to the laser peak. Therefore, the possibility of controlling the laser intensities with the electrooptic effect using electric fields well below the coercive field of the ferroelectric material could depend to a great extent on the linewidth of the laser line to be controlled. In this sense, laser transitions of $Nd^{3+}$ ions in YAG with 0.5 nm linewidth or in $YVO_4$ with 0.8 nm linewidth [32] could be not optimum for pumping if one wants to use the linear electrooptic effect to control laser intensity. Instead, laser emissions of materials with larger fluorescence lifetimes giving smaller laser spectral linewidths, like Nd-doped $YLiF_4$ ($\approx$ 0.2 nm linewidth) are more desirable. In this way, we will need only to shift the QPM peak by 0.1 nm approximately to obtain zero intensity of the laser lines in the ideal case, thus reducing considerably the electric fields needed.

## 3. Conclusions

We have modelled the working of quasi-periodic poling ferroelectric non-linear crystals LN and SBN under QPM condition for intracavity conversion. We have show a fabrication scheme in order to control the RGB laser intensities in an independent way by means of the linear electrooptic effect. LN seems to be not suitable for electrooptic control of laser intensities according to our design principle due to the relatively small electrooptic coefficients and the control of laser intensities by varying the phase mismatch of the input interacting waves seems to be more convenient. However, SBN can be a very important material for optoelectronic applications because electrooptic control of laser emission intensities can be performed in this material according to the simple designed principle of Fig. 2. The electrooptic control can be achieved in SBN due to its outstandingly large electrooptic coefficients if the RGB laser linewidths are small enough and the periodic poling errors are kept under control below certain limits.

**Acknowledgments:** The authors thank J. Capmany and A. Segura for stimulating discussions.

# Figure captions

**Fig. 1.** Schematic show of the $Nd^{3+}$ crystal-field levels in Nd:APLN and Nd:YVO$_4$ and the pump and infrared laser lights used for simultaneous generation of the three fundamental colors in Nd:APLN and SBN60.

**Fig. 2.** Schematic show of the high-brightness display suggested for simultaneous generation of the three fundamental colors [red (R), green (G), and blue (B)] with independent control of laser intensities by means of the electrooptic effect. The axis of light propagation is the x axis and the direction of application of the electric field is the z axis. The inset shows the composition of the aperiodic poling units.

**Fig. 3.** Position of the QPM peaks in the wavevector representation in quasi-periodically poled LiNbO$_3$ in agreement with the laser peaks observed in Nd:APLN (a) and its correlation with the crystal zone contributing to each laser color through the total length of the crystal (b).

**Fig. 4.** Fourier Transform calculated efficiency of the SFD and SSFM processes in the visible range in LN as a function of the asymmetry parameter using the laser frequencies of Ref. [6].

**Fig. 5.** Working procedure of the electric field for controlling the laser intensity. (a) If no electric field is applied, the laser light intensity is maximum because the QPM peak is tunned to the laser peak. (b) With a small applied electric field the laser light intensity decreases because of the detuning of the QPM and laser peaks. (c) If a large enough electric field is applied the total detuning makes the laser intensity fall to zero.

**Fig. 6.** Calculated electric fields needed to shift the QPM peak by 0.2 nm from the different laser wavelengths observed in SBN60 as a function of the asymmetry parameter. Red (solid line), green (dashed line), and blue (dotted line). Also shown are the calculated electric fields that would be needed to shift the same colors in SBN75 with a larger electrooptic coefficient.

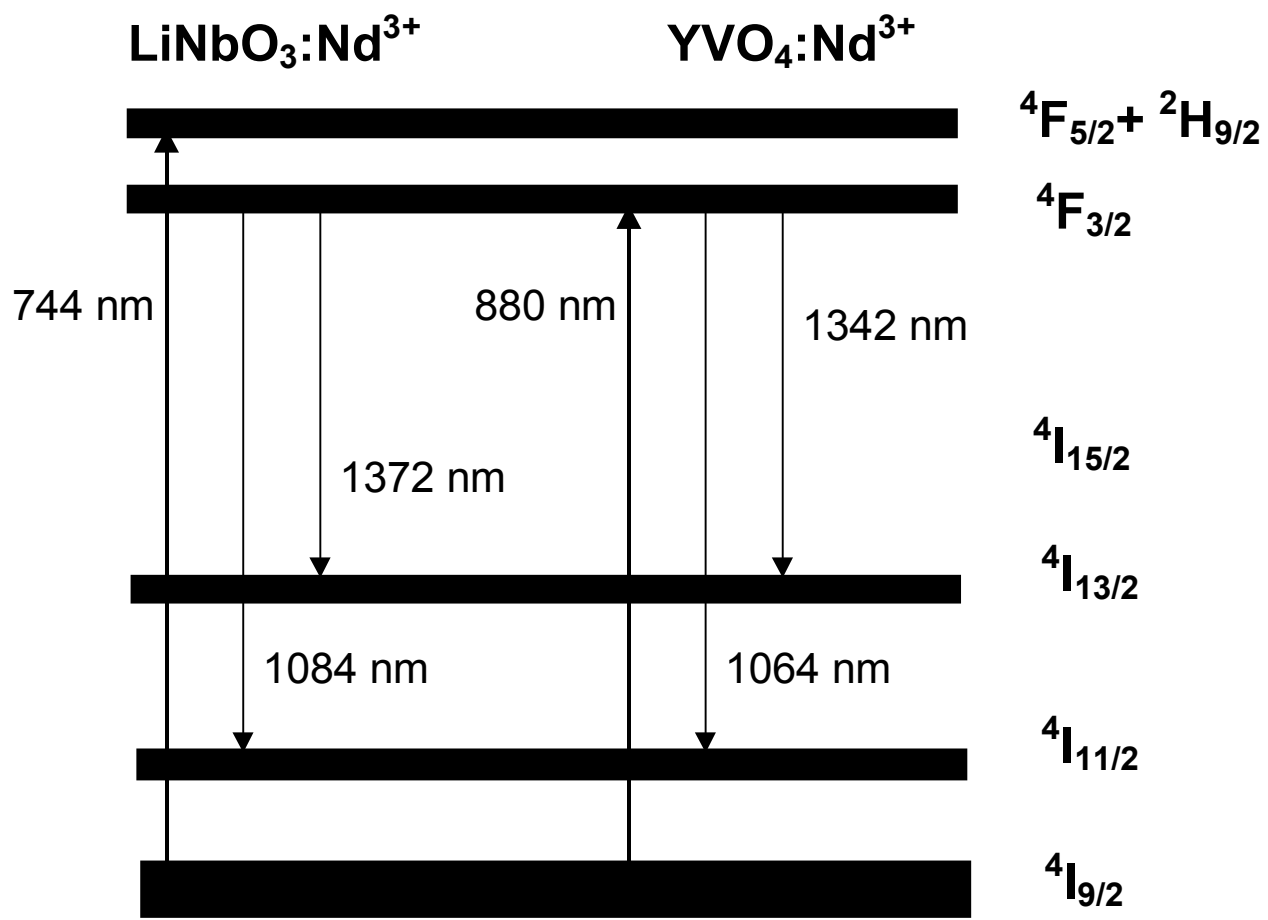

Figure 1. R. Vilaplana et al.

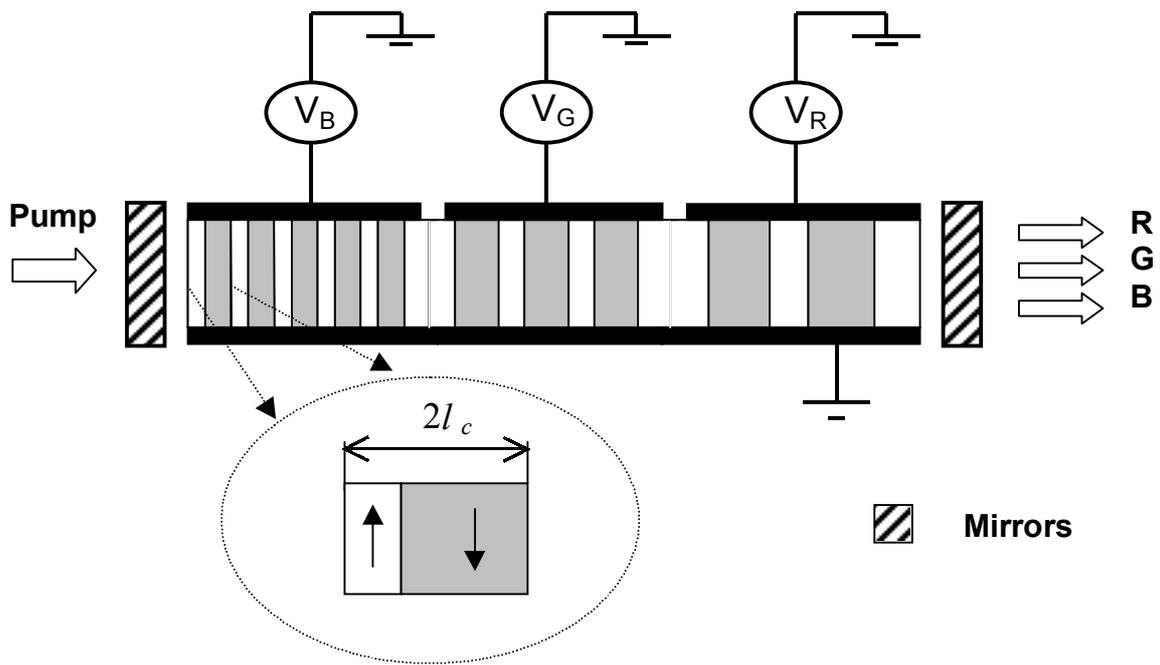

Figure 2. R. Vilaplana et al.

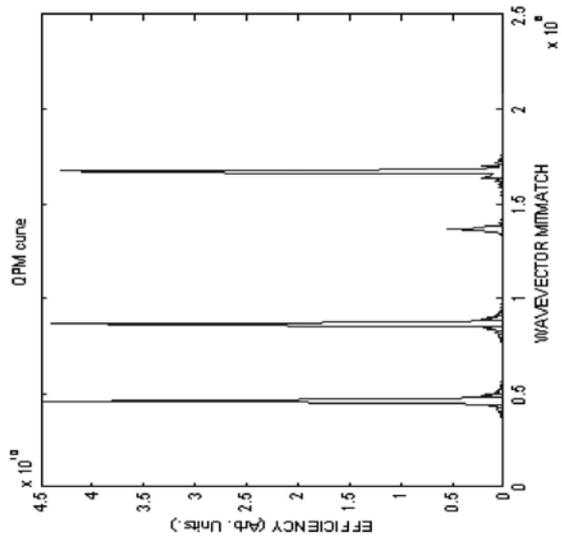
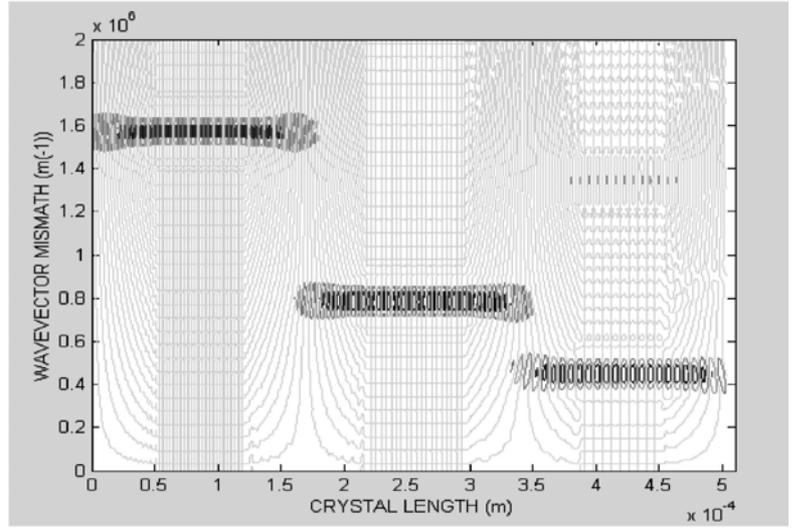
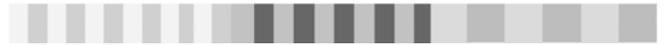

Figure 3. R. Vilaplana et al.

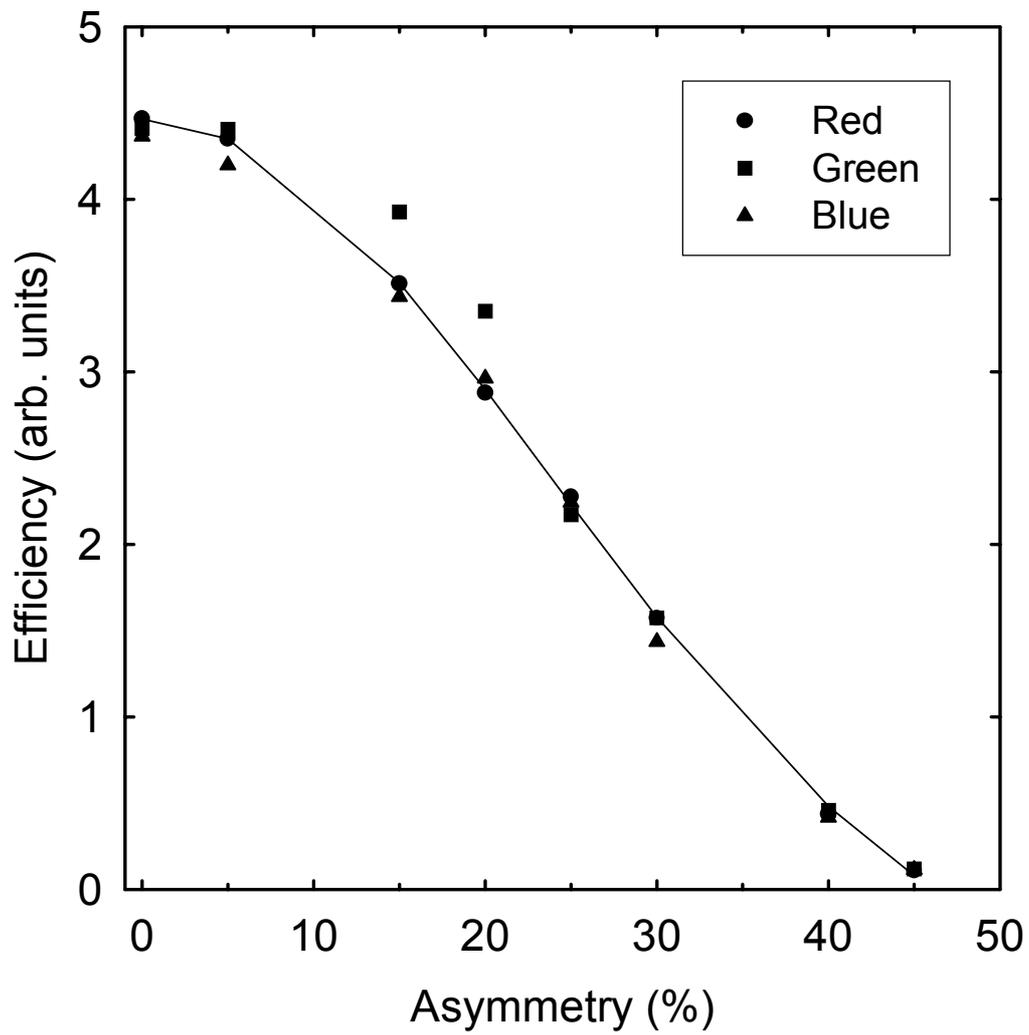

Figure 4. R. Vilaplana et al.

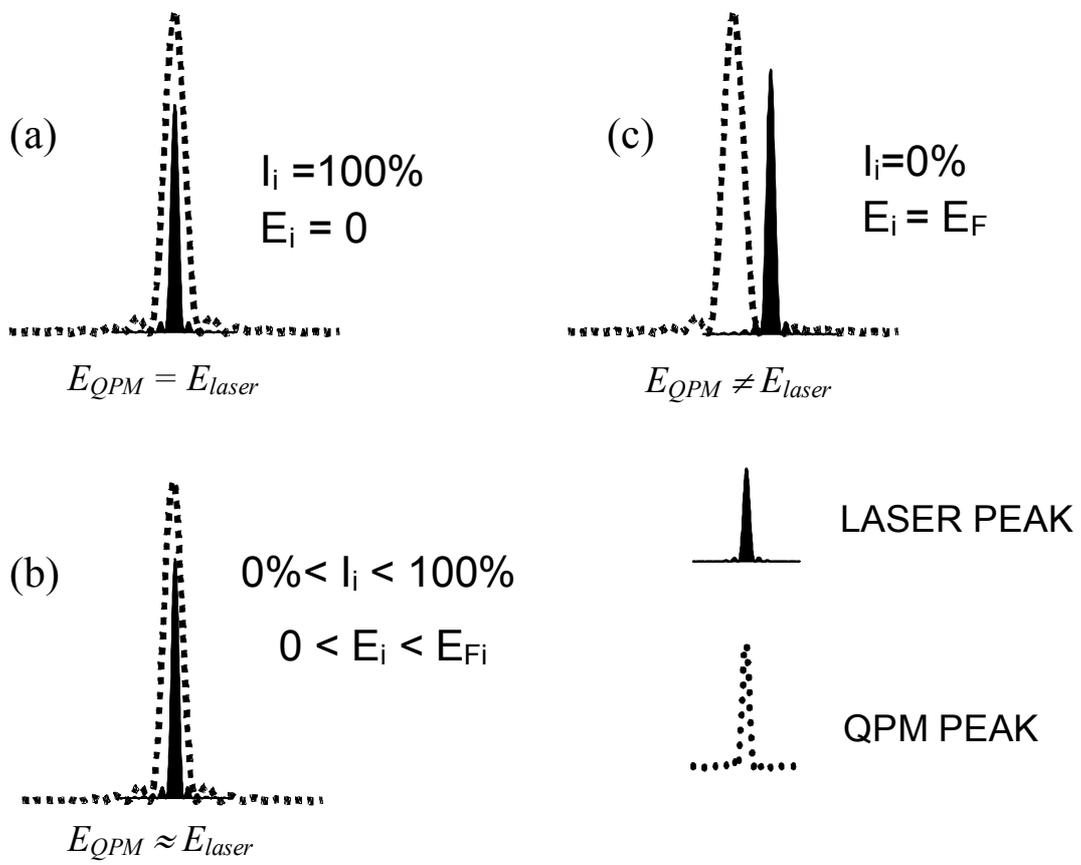

Figure 5. R. Vilaplana et al.

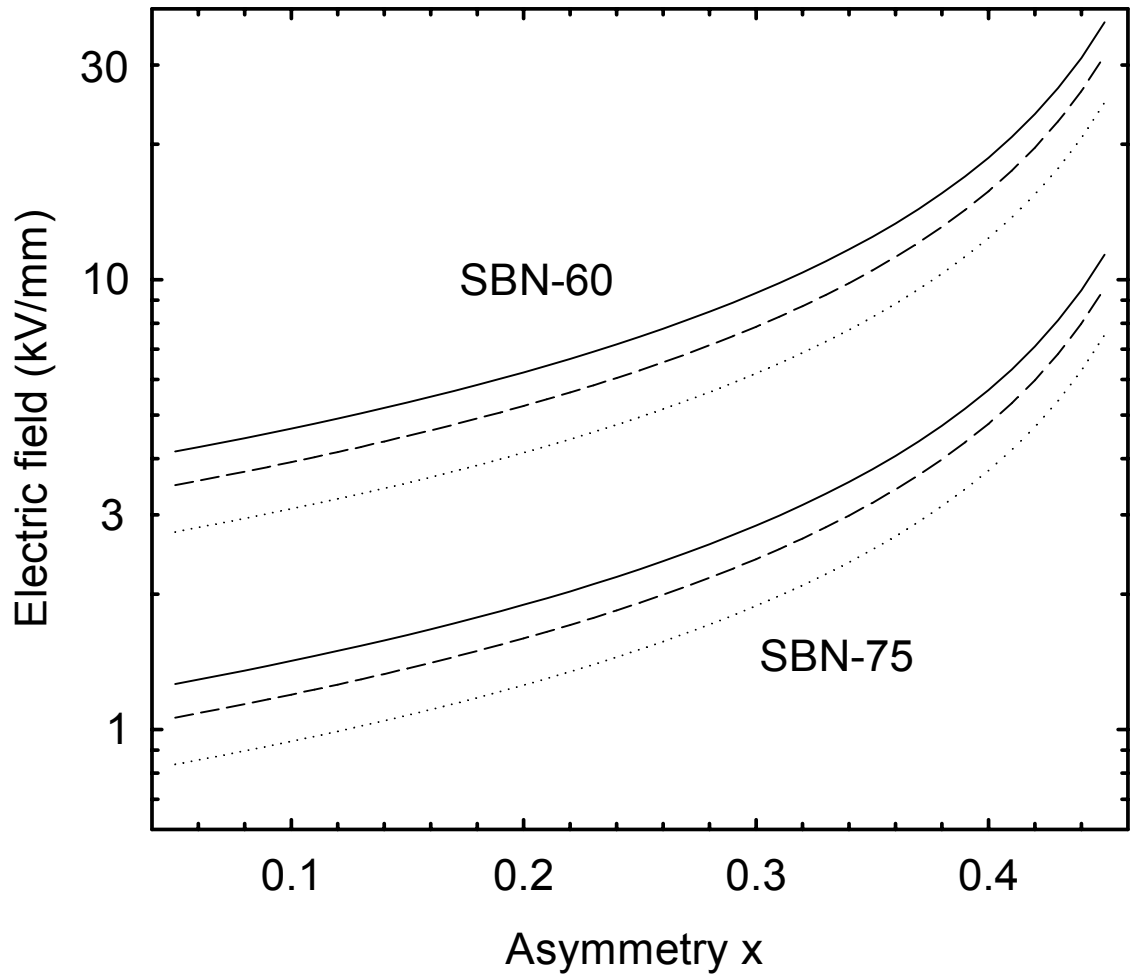

Figure 6. R. Vilaplana et al.